\documentclass[sigconf,nonacm]{acmart}
\AtBeginDocument{%
  }

\setcopyright{none}

\begin{document}

\title{Probabilistic Analysis of Copyright Disputes and Generative AI Safety}

\renewcommand{\footnotemark}{}

\author{Hiroaki Chiba-Okabe}
\authornotemark[]
\affiliation{%
  \institution{University of Pennsylvania}
  \city{Philadelphia}
  \state{Pennsylvania}
  \country{USA}
}

\renewcommand{\shortauthors}{Chiba-Okabe}
\newcommand{\indep}{\perp \!\!\! \perp}

\begin{abstract}
This paper presents a probabilistic approach to analyzing copyright infringement disputes. Evidentiary principles shaped by case law are formalized in probabilistic terms, and the ``inverse ratio rule''---a controversial legal doctrine adopted by some courts---is examined. Although this rule has faced significant criticism, a formal proof demonstrates its validity, provided it is properly defined. The probabilistic approach is further employed to study the copyright safety of generative AI. Specifically, the Near Access-Free (NAF) condition, previously proposed as a strategy for mitigating the heightened copyright infringement risks of generative AI, is evaluated. The analysis reveals limitations in its justifiability and efficacy.
\end{abstract}




\settopmatter{printacmref=false, printccs=false, printfolios=false}
\maketitle

\section{Introduction}
Copyright law grants creators exclusive rights to their original works across diverse categories of creative expression, such as literature, music, and art, with the aim of incentivizing creation and enriching society \cite{gorman2006copyright}. The rise of new technologies has continuously challenged conventional copyright principles, and in response, copyright jurisprudence has evolved over time, becoming increasingly complex \cite{joyce2020}. Despite the ever-growing need to understand how copyright law interacts with emerging technologies---generative AI being a prominent recent example \cite{samuelson2023,lemley2024}---the theoretical framework remains underdeveloped, leading to significant legal uncertainty \cite{depoorter2009}.

This article employs a probabilistic approach to analyze copyright infringement disputes, seeking to bring clarity to its complexities. Based on the view that the legal fact-finding process can be understood to involve the evaluation of the probabilities that the allegations are true or false based on available evidence \cite{cullison1969,finkelstein1970,stein2005}, the core evidentiary principles of copyright jurisprudence are formalized in probabilistic terms. This allows for a mathematical analysis of a controversial copyright legal doctrine, the ``inverse ratio rule,'' whose validity has been the subject of considerable debate \cite{aronoff2007exploding,balganesh2022}. While a previous study \cite{balganesh2022} notes a link between the probabilistic nature of legal fact‑finding and the inverse ratio rule, no formal probabilistic treatment of this rule appears in the literature. This paper fills that gap by offering a formal analysis of the inverse ratio rule, demonstrating that the rule must hold under natural assumptions when its content and scope are carefully delineated. 

The same approach is further utilized to evaluate the Near Access-Free (NAF) condition \cite{vyas2023}, which has been proposed to mitigate the copyright infringement risks posed by generative AI. While the authors who introduced the NAF condition left its precise legal characterization an open question \cite{vyas2023}, subsequent studies have examined the NAF condition from a legal perspective \cite{lee2024talkin,elkinkoren2024}. This paper advances the discussion by revealing that, in certain contexts, the NAF condition operates in an ethically questionable way---reducing transparency about the materials generative AI accessed during training---and that its effectiveness remains limited.

\section{Probabilistic Analysis of Copyright Disputes}

\subsection{Formalization of Evidentiary Principles}

To prevail in a copyright infringement dispute, the copyright owner, as the plaintiff, must demonstrate two elements under the traditional \emph{Arnstein} test: ``(a) that defendant copied from plaintiff's copyrighted work and (b) that the copying (assuming it to be proved) went so far as to constitute improper appropriation.''\footnote{\textit{Arnstein v. Porter}, 154 F.2d 464, 468 (2d Cir. 1946).} These two requirements are often referred to simply as ``factual copying'' and ``improper appropriation,'' respectively \cite{balganesh2022}.

Direct evidence of factual copying includes admission by the defendant or an eyewitness account of the replication or appropriative behavior, but such evidence is hard to obtain.\footnote{See, e.g., \textit{Smith v. Jackson}, 84 F.3d 1213, 1218 (9th Cir. 1996), stating ``direct evidence of copying is not available in most cases.''} Therefore, factual copying is almost invariably proved by circumstantial evidence \cite{gorman2006copyright,balganesh2022}, and the evidence generally accepted by courts consists of that pertaining to access and similarity. Specifically, copying is proved based on (i) evidence showing that the defendant had access to the original work and (ii) the similarity between the original work and the defendant's work \cite{balganesh2022,loren2019}. Courts generally hold that proof of access combined with a certain level of similarity, that is probative of copying, is sufficient to establish factual copying \cite{latman1990,loren2019}. Access pertains to the antecedent of factual copying, in the sense that the defendant having had access to the original work is a necessary condition for engaging in copying \cite{balganesh2022,latman1990}. In contrast, similarity emerges as a result of copying \cite{balganesh2022,latman1990}. When the similarity between the two works is so pronounced as to be deemed ``striking,'' this alone suffices to establish that copying has occurred.\footnote{\textit{Arnstein}, 154 F.2d at 468-69.} In such cases, it can be inferred that the defendant had access to the original work based on the high degree of similarity.\footnote{\textit{Ty, Inc. v. GMA Accessories, Inc.}, 132 F.3d 1167, 1170 (7th Cir. 1997), stating ``a similarity that is so close as to be highly unlikely to have been an accident of independent creation is evidence of access.''}

The second requirement, improper appropriation, is established by demonstrating that the extent of copying reaches the threshold at which the similarity between the original work and the defendant's work can be deemed ``substantial'' in degree. The determination of this requirement involves a normative judgment, taking into account both quantitative and qualitative aspects of copying \cite{gorman2006copyright,balganesh2012}. Thus, similarity serves a dual role: It supports the finding of factual copying \emph{and} acts as an independent criterion for establishing improper appropriation \cite{latman1990,balganesh2022}, though the sense of similarity differs in each context, even if there may be some overlap \cite{loren2019}.

In court proceedings, adjudicators draw on knowledge of the frequencies of occurrences of certain events and patterns derived from experience to evaluate the likelihood of the particular circumstances in question \cite{stein2005}. Although the absolute truth typically cannot be attained, the final decision in most civil cases is made based on the ``preponderance of evidence.'' This standard requires the party bearing the burden of proof to demonstrate that their claim is more likely true than not. From a probabilistic perspective, it is often understood as requiring a demonstration of a likelihood greater than 50\% that the claim is true \cite{mccauliff1982,stein2005}, particularly in binary situations where the question is whether an assertion is true or false \cite{cheng2013}. 

Building on this view and considering a discrete probability space $(\Omega,\mathcal{F},P)$, essential copyright evidentiary principles can be formalized in probabilistic terms. The condition for copyright infringement to exist can be expressed using random variables as $Copy=1 \wedge Substantial=1$. Here, $Copy$ is a binary random variable representing the presence ($Copy=1$) or absence ($Copy=0$) of factual copying of the plaintiff's copyrighted work during the creation of the defendant's work. Similarly, $Substantial$ represents the presence or absence of substantial similarity between the two works. Typically, adjudicators can determine whether substantial similarity exists, or the truth value of $Substantial$, with certainty by directly comparing the original work to the defendant's work presented at trial. In contrast, the absolute truth about whether the defendant engaged in factual copying remains unknown during the trial, and the plaintiff must present sufficient evidence for the adjudicators to conclude that the probability of $Copy=1$ exceeds a certain threshold, $\lambda$, corresponding to the preponderance of evidence standard.

The case law establishing that probative similarity combined with access typically suffices to prove copying can be represented in probabilistic terms as $P(Copy=1|Access=1,\,Probative =1) = \eta$ where $\eta > \lambda$. $Access$ and $Probative$ are binary random variables representing whether access by the defendant to the plaintiff's original work was present and whether the original and defendant's works are probatively similar, respectively. Suppose the adjudicators learn from the evidence presented during the trial that the defendant indeed had access to the original work ($Access=1$) and that the defendant's work is probatively similar to the original work ($Probative=1$). This results in the adjudicators' degree of belief that copying occurred ($Copy=1$) surpassing the threshold $\lambda$ (since $\eta>\lambda$), thus meeting the standard of proof.

This approach can accommodate the aforementioned logical relationship between factual copying and access, which can be represented as $P(Copy=1|Access=0)=0$. A further step can be taken to represent the evidentiary rule that striking similarity is typically sufficient to establish copying by stipulating $P(Copy=1|Access=1, Striking=1)=\rho$ and $P(Access=1|Striking=1)=\sigma$ where $\rho\sigma>\lambda$. $Striking$ is a binary random variable corresponding to the presence or absence of striking similarity. Assuming that the adjudicators learn during the trial that $Striking=1$ with certainty, taking into account the logical relationship between access and copying, the degree of belief about copying exceeds the threshold, as desired: $P\left(Copy=1|Striking=1\right)=P\left(Copy=1|Striking=1,Access=1\right) P\left(Access=1|Striking=1\right)+0\cdot P\left(Access=0|Striking=1\right)=\rho\sigma >\lambda$.

\subsection{The Inverse Ratio Rule}

According to the inverse ratio rule, a stronger proof of access lowers the level of similarity required to establish copyright infringement, while a higher level of similarity reduces the burden of proof regarding access \cite{aronoff2007exploding,balganesh2022}. Although the rule has been applied for nearly a century, serious objections have been raised to its application \cite{aronoff2007exploding}. The controversial status of the inverse ratio rule gained renewed attention with its recent abrogation by the Ninth Circuit Court.\footnote{\textit{Skidmore v. Led Zeppelin}, 952 F.3d 1051, 1066-69 (9th Cir. 2020) (en banc).}

Some courts have applied the inverse ratio rule to ease the substantial similarity threshold for proving improper appropriation when access is strongly established \cite{balganesh2022,aronoff2007exploding}, but such application of the rule lacks clear justification \cite{balganesh2022}. A more promising formulation of the inverse ratio rule concerns the factual copying prong: Strong evidence of access lowers the similarity needed to prove factual copying, while weak evidence of access requires greater similarity; conversely, a high degree of similarity can compensate for limited evidence of access \cite{balganesh2022}. In fact, this version of the inverse ratio rule can be \emph{proved} by considering a set of random variables (Definition~\ref{def:variables}) that encapsulates the evidentiary principles discussed in the preceding section in a more generalized form.

\begin{definition}\label{def:variables} $C,A\in\{0,1\}$, $S\in\{1,2,\cdots,n\}$, and $EA\in\{1,2,\cdots,m\}$ are random variables that satisfy:
\begin{align}
&P(C=1|A=0)=0\label{def:variables_logic}\\
&P(C=1|A=1,\, S=i)=\alpha_{i},\, \forall i\label{def:variables_alpha}\\
&P(A=1|S=i ,\, EA = j)= \beta_{i,j},\,\forall (i,j)\label{def:variables_beta}\\
&EA\indep \{C,S\}|A \label{def:variables_indep}
\end{align}
where $\alpha_{i+1}\geq\alpha_{i}$ for all $i$, $\beta_{i,j+1} \geq \beta_{i,j}$ for all $j$ with $i$ fixed, and $\beta_{i+1,j} \geq \beta_{i,j}$ for all $i$ with $j$ fixed.
\end{definition}
$C$ and $A$ are abbreviations of $Copy$ and $Access$, respectively, from earlier discussions, which enjoy a logical relationship (Definition~\ref{def:variables} (\ref{def:variables_logic})). The value of $S$ represents the level of similarity in the probative sense, which serves as evidence to prove factual copying (Definition~\ref{def:variables} (\ref{def:variables_alpha})). It is reasonable, at least in principle (assuming that no other specific circumstances suggest otherwise), to think that there is a monotonic relationship among the inferential values of different degrees of similarity; that is, a higher degree of similarity more strongly infers factual copying, so $\alpha_{i+1} \geq \alpha_{i}$ holds. Similarity also has evidentiary value in establishing access to the original work (Definition~\ref{def:variables} (\ref{def:variables_indep})), as demonstrated by the evidentiary rule regarding striking similarity. It is natural to posit that such evidentiary value for access increases monotonically with similarity in principle, leading to the assumption that $\beta_{i+1,j} \geq \beta_{i,j}$ for all $i$.

Variable $EA$ is introduced to capture the interaction between similarity and evidence of access (other than similarity itself). The values it can take correspond to varying degrees of the strength of evidence of access presented in court (Definition~\ref{def:variables} (\ref{def:variables_beta})). For example, strong evidence of access might include testimony that the original work was directly handed to the defendant or the fact that the work was publicly accessible and widely known. Conversely, the fact that the original work was kept private, with little chance of the defendant encountering it, would imply a low likelihood of access. Such evidence is assumed to directly impact only the inference of access as specified in Definition~\ref{def:variables} (\ref{def:variables_indep}), which states that $C$ and $S$ are independent of $EA$ conditional on $A$. The differing strengths of evidence naturally create an ordered variation in the inferential value, captured by the relationship $\beta_{i,j+1} \geq \beta_{i,j}$ for all $j$.

During court proceedings, adjudicators learn both the level of similarity between the works and the strength of evidence of access presented, corresponding to a pair of values $(i,j)$. Then, the degree of belief of the adjudicators about the presence of factual copying, $C=1$, is formed given the knowledge of $S=i$ and $EA=j$. 

Under these assumptions, the inverse ratio rule as it pertains to factual copying can now be precisely articulated in formal language and its correctness proved, as summarized in Proposition~\ref{prop:irr} (see Appendix~\ref{appendix:irr_strong} for a stronger version of the inverse ratio rule that holds with strict inequalities). Proofs are shown in the Appendix.
\begin{proposition} The following inverse ratio rule holds for any $\lambda\in(0,1)$: If $j \geq j^{\prime}$, then $\min\{i:P(C=1|S=i,\, EA=j)> \lambda\} \leq \min\{i:P(C=1|S=i,\, EA=j^{\prime})> \lambda\}$. Similarly, if $i\geq i^{\prime}$, then $\min\{j:P(C=1|S=i,\, EA=j)> \lambda\} \leq \min\{j:P(C=1|S=i^{\prime},\, EA=j)>\lambda\}$.
\label{prop:irr}
\end{proposition}

Even though access is merely a necessary condition for copying, the \textit{strength} of its evidence still affects the inference of copying by lowering the required similarity, despite critics’ argument that the inverse ratio rule improperly lowers the similarity threshold \cite{aronoff2007exploding}. This is because the likelihood of copying is the product of two factors, (i) the likelihood of copying given the presence of access and the level of similarity and (ii) the likelihood of access given the level of similarity and other evidence of access, and stronger (weaker) evidence of access enlarges (dwindles) the second factor. 

Furthermore, the justification offered by the proponents of the inverse ratio rule is clarified in this paper; it is argued that moderate evidence of access and similarity can jointly support an inference of copying \cite{balganesh2022}. However, since the product of two probabilities is no greater than either individual factor, combining small factors, stemming from weak evidence in both similarity and access, cannot yield a stronger inference. Instead, similarity and access work complementarily rather than cumulatively: Stronger evidence of access raises factor (ii), and even when factor (i) is modest due to a somewhat low level of similarity, their product can still clear the threshold for factual copying, and vice versa.

\section{Near Access-Freeness as a Framework for Copyright Safety of Generative AI}\label{sec:genai}

When generative AI is used to create content, it is typically offered as a service by the AI developer and used by a user. In such cases, both the user and the AI developer may be considered direct or indirect infringers if the generated output is substantially similar to an original work, depending on the specific circumstances. Copyright infringement may occur even without the user or developer intending to replicate the copyrighted material or to create something similar \cite{lee2024talkin}. Due to the uninterpretability of generative models, it can be challenging to provide direct evidence of factual copying for AI-generated content. In most cases, access and similarity will likely serve as the primary evidence of infringement, much like in traditional copyright disputes \cite{lee2024talkin}. The inclusion of a work in a model’s training corpus is often viewed as sufficient evidence of the access element required to establish copying \cite{lee2024talkin,elkinkoren2024}. 

Near Access-Freeness \cite{vyas2023}, which aims to mitigate copyright infringement risks, ensures that the generation probability of a generative model is closely aligned with that of a version of the model that has not had access to copyrighted data. In a canonical form \cite{vyas2023,elkinkoren2024}, the NAF condition can be understood to impose the inequality $P(Z=z|Access=1)\leq e^{\epsilon}P(Z=z|Access=0)$
for any $z$, where $Z$ is a random variable representing the output of the generative model, $Access$ indicates whether the model had access to a copyrighted work, and parameter $\epsilon>0$ determines the strength of theoretical guarantee. The output distribution of the AI model must satisfy this inequality with respect to any copyrighted work in the training data. In words, if a model meets the NAF condition with some $\epsilon$, having access to the original work during training can raise the probability of generating any particular output by at most a factor of $e^{\epsilon}$. For instance, a model trained on Snoopy images is at most $e^{\epsilon}$ times more likely to produce any image---including one that resembles Snoopy---than a version of the model that never had access to Snoopy images in training. The NAF condition with a specific parameter value $\epsilon$ will be referred to as $\epsilon$-NAF herein. 


The authors who proposed the NAF condition leave open the question of whether this approach could function as a valid defense in court \cite{vyas2023}. Although the NAF condition has been acknowledged to be at least partially effective in reducing the risks of infringement \cite{elkinkoren2024}, it has also faced skepticism for not fully aligning with the principles of copyright law \cite{lee2024talkin}. Here, a probabilistic formalism is employed as an aid for precise analysis to reconcile divergent views.

One straightforward way the NAF condition contributes to copyright safety is by addressing the \emph{prospective} problem of how likely the model is to generate a copyright-infringing output. It can be concluded that this probability is low, given that (i) the model satisfies $\epsilon$-NAF with a small $\epsilon$ and (ii) the model is unlikely to produce outputs highly similar to copyrighted works if it had been trained without access. This trivially follows from standard probabilistic reasoning by viewing the substantial similarity of a generated work to a fixed copyrighted work as a function of the generated work. Specifically, letting $Substantial(Z)$ be a binary function of random variable $Z$ that takes value 1 if the similarity of an output $z$ to the copyrighted work reaches the threshold corresponding to substantial similarity, $\epsilon$-NAF ensures that: $P(Substantial(Z)=1|Access=1)\leq e^{\epsilon} P(Substantial(Z)=1|Access=0)$. Due to this inequality, given that $P(Substantial(Z)=1|Access=0)$ is small and $e^{\epsilon}$ is close to 1, corresponding to a small value of $\epsilon$, $P(Substantial(Z)=1|Access=1)$ must also be small. Consequently, the model is unlikely to produce substantially similar content or infringe copyright, as it would not meet the improper appropriation prong.

However, the \emph{retrospective} problem of determining whether a specific generated output infringes copyright---the issue adjudicators face in court---is more nuanced. Drawing a conclusion based on broad statistical data that generative models satisfying the NAF condition produce substantially similar outputs with low probability, without assessing case-specific details pertaining to the substantial similarity between the specific output and the original work at hand, constitutes a fallacy that must be avoided \cite{tribe1971,allen2013}. The mere rarity of substantially similar outputs generated by such models does not influence the evaluation of substantial similarity between the two specific works, which adjudicators can typically ascertain with certainty by directly examining both works presented in court. Therefore, linking the NAF condition to the improper appropriation prong in this context appears unfounded.

The focus now shifts to examining whether and how the NAF condition affects the determination of the other prong, factual copying, within the context of the retrospective problem. On one hand, there seems to be little justification for expecting the NAF condition to meaningfully influence the finding of copying, \emph{given} both the presence of access and a fixed level of similarity in the probative sense. Specifically, the theoretical guarantee of the NAF condition does not directly impose constraints on factual copying; instead, it merely dictates a probabilistic relationship between access and generated outputs. Furthermore, as observed in \cite{lee2024talkin}, the assertion that a similar output could have been produced by copying from another source does not constitute a valid defense under copyright law. 

On the other hand, the NAF condition can impact the inference of access based on similarity, as it influences the observable distribution of similarity between generated content and copyrighted works. This becomes evident through a formal analysis employing the random variables as defined in Definition~\ref{def:variables_naf}.
\begin{definition}\label{def:variables_naf} $A_{M}\in\{0,1\}$ and $EA_{M}\in\{1,2,\cdots,m\}$ are random variables that satisfy:%
\begin{align}
&P(S=i|A_{M}=1)\leq e^{\epsilon}P(S=i|A_{M}=0),\,\forall i\tag{1}\label{def:variables_naf_epsilon}\\
&P(A_{M}=1|EA_{M}=k)= \delta_{k},\forall k\tag{2}\label{def:variables_naf_delta}\\
&EA_{M}\indep S|A_{M}\tag{3}\label{def:variables_naf_indep}
\end{align}%
where $\delta_{k+1}\geq\delta_{k}$.
\end{definition}
In the context of AI generation, access can refer to access by the user or access by the AI model. To clarify that access by the model is at issue, the variable for access, $A_{M}$, and the corresponding evidence, $EA_{M}$, are denoted with the subscript $M$. The NAF condition (Definition~\ref{def:variables_naf} (\ref{def:variables_naf_epsilon})) and some degree of belief, $\delta_{k}$, about the likelihood of access by the AI model given available evidence (other than similarity) of access (Definition~\ref{def:variables_naf} (\ref{def:variables_naf_delta})) are assumed. Analogous to $EA$ in Definition~\ref{def:variables}, $EA_{M}$ directly impacts only the inference of access (Definition~\ref{def:variables_naf} (\ref{def:variables_naf_indep})).

\begin{proposition}
For any $\epsilon>0$ and $(i,k)$, $P(A_{M}=1|S=i,\, EA_{M}=k)\leq  \Gamma(\epsilon,\delta_{k})$ where $\Gamma(\epsilon, \delta):=\frac{e^{\epsilon} \delta}{1+\delta(e^{\epsilon}-1)}$. $\Gamma(\epsilon,\delta)$ is increasing in $\epsilon$ and $\delta$, and satisfies $\delta\leq \Gamma(\epsilon,\delta)$ and $\lim\limits_{\epsilon\rightarrow 0}\Gamma(\epsilon,\delta)=\delta$.
\label{prop:naf}
\end{proposition}
Proposition~\ref{prop:naf} shows that the NAF condition indeed imposes an upper bound on the probability of finding the presence of access to the original work by the model. A lower value of $\epsilon$ enhances the effectiveness of this upper bound, and, in the limit of $\epsilon\rightarrow 0$, the evidentiary value of similarity in inferring access vanishes. In other words, similarity fails to serve as meaningful additional evidence of access, with the value of $P(A_{M}=1|S=i,\, EA_{M}=k)$ converging to $\delta_{k}=P(A_{M}=1|EA_{M}=k)$ for all $i$. 

However, the way in which the NAF condition operates in this context is ethically debatable. Intuitively, when the NAF condition is satisfied with small $\epsilon$, training with access to the original work does not increase the model’s probability of generating any particular output much (not more than a multiplicative factor $e^{\epsilon}$) compared to training without access. This, in turn, ensures that the model is not significantly more likely to produce outputs similar to the original work relative to a version of the model that did not have access. Hence, one cannot strongly infer that the model must have been trained on the original work just because it produces a similar output. However, this approach may contradict the demand for transparency in training data, as advocated by researchers \cite{sag2023,Yoo2024}. Furthermore, $\Gamma$ is bounded below by the degree of belief $\delta_{k}$ that the model had access to the specific original work in question, given available evidence of access other than similarity. The upper bound on the degree of belief about access becomes less significant as the baseline belief $\delta_{k}$ increases, either through stronger evidence of access to the original work by the model (e.g., unequivocal evidence of its inclusion in the training data) or through the increasingly common practice of training AI models on broader sets of copyrighted materials.

\section{Concluding Remarks}

Probabilistic analysis has been shown to enhance the understanding of copyright jurisprudence and its interaction with generative AI. However, there are certain limitations to the analysis presented in this paper. While the analysis is based on a probabilistic view of legal fact-finding, this perspective has faced sustained criticism \cite{tribe1971,allen2013}; evaluating the validity of probabilistic approaches in legal fact-finding in general falls beyond the scope of this paper. Furthermore, the analysis abstracts from certain details involved in AI generation, such as the content of user-provided prompts, which may influence the identification of direct or indirect infringers \cite{lee2024talkin}.

Several future directions remain to be explored. Firstly, the presented approach has wider potential applications within copyright jurisprudence, where many legal doctrines can be understood from evidentiary perspectives \cite{lichtman2003}. In addition, a variety of formal and computational methodologies \cite{verheij2015,halpern2017} for evidentiary and probabilistic reasoning beyond the simple framework used here could be applied to further refine and operationalize the probabilistic approach. Lastly, future research should analyze and develop copyright risk mitigation strategies alternative to the NAF condition.

\begin{acks}
The author acknowledges financial support from the Simons Foundation Math+X Grant to the University of Pennsylvania.
\end{acks}

\bibliographystyle{ACM-Reference-Format}
\bibliography{sample-base}

\appendix

\section*{Appendix}

\section{The Inverse Ratio Rule with Strict Inequality}\label{appendix:irr_strong}

The inverse ratio rule holds with strict inequalities under moderate additional assumptions (Proposition~\ref{prop:irr_strong}). 

\begin{proposition}\label{prop:irr_strong} The following strong inverse ratio rule holds for any $\lambda\in(0,1)$: Suppose $j>j'$ and $\mu':=\min\{i:P(C=1|S=i,\, EA=j')>\lambda\}>1$. If $\tfrac{\alpha_{\mu'}}{\alpha_{\mu'-1}} \leq \tfrac{\beta_{\mu'-1,j}}{\beta_{\mu',j'}}$, then $\min\{i:P(C=1|S=i,\, EA=j)>\lambda\}<\mu'$. Similarly, suppose $i>i'$ and $\kappa':=\min\{j:P(C=1|S=i',\, EA=j)>\lambda\}>1$. If $\tfrac{\beta_{i',\kappa'}}{\beta_{i,\kappa'-1}} \leq \tfrac{\alpha_{i}}{\alpha_{i'}}$, then $\min\{j:P(C=1|S=i,\, EA=j)>\lambda\}<\kappa'$.
\end{proposition}

\section{Proofs}\label{appendix:proof}
\begin{proof}[Proof of Proposition~\ref{prop:irr}]
From the logical relationship between copying and access as well as $EA$'s conditional independence, we have $P\left(C=1|S=i,\, EA=j\right)=\alpha_{i}\beta_{i,j}$ for any $(i,j)$.

The first element of the inverse ratio rule is proved by showing contradiction. Assume $\mu > \mu^{\prime}$ where $\mu:=\min \{i:P(C=1|S=i,\, EA=j)>\lambda\}$ and 
$\mu^{\prime}:=\min \{i:P(C=1|S=i,\, EA=j^{\prime})>\lambda\}$.

Then, by assumptions about the ordering of the terms of the form $\alpha_{i}$ and $\beta_{i,j}$, the inequalities $\lambda <\alpha_{\mu^{\prime}} \beta_{\mu^{\prime},j^{\prime}}\leq \alpha_{\mu^{\prime}} \beta_{\mu^{\prime},j}$ hold. However, this implies that $\mu^{\prime}$, which is strictly smaller than $\mu$, would have achieved $P(C=1|S=\mu^{\prime},\, EA=j)>\lambda$, contradicting the assumption that $\mu$ was the minimum that achieves this. 

The proof for the second element of the inverse ratio rule is similar. Assume $\kappa>\kappa^{\prime}$ where $\kappa:=\min \{j:P(C=1|S=i,\, EA=j)>\lambda\}$ and $\kappa^{\prime}:=\min \{j:P(C=1|S=i^{\prime},\, EA=j)>\lambda\}$. Then, the inequalities $\lambda <\alpha_{i^{\prime}}\beta_{i^{\prime},\kappa^{\prime}}\leq \alpha_{i} \beta_{i,\kappa^{\prime}}$ hold, which is in contradiction with the assumption of minimality of $\kappa$.
\end{proof}

\begin{proof}[Proof of Proposition~\ref{prop:naf}]
By the NAF condition and the conditional independence of $EA_{M}$:
\begin{displaymath}
\begin{split}
&P(A_{M}=1|S=i,\, EA_{M}=k)\\
&=\frac{P(S=i|A_{M}=1,\, EA_{M}=k)P(A_{M}=1| EA_{M}=k)}{P(S=i| EA_{M}=k)}\\
&\leq \frac{e^{\epsilon}P(S=i|A_{M}=0,\, EA_{M}=k)P(A_{M}=1|EA_{M}=k)}{P(S=i| EA_{M}=k)}\\
&=\frac{e^{\epsilon}\frac{P(A_{M}=0|S=i,\, EA_{M}=k)P(S=i| EA_{M}=k)}{P(A_{M}=0|EA_{M}=k)}P(A_{M}=1| EA_{M}=k)}{P(S=i|EA_{M}=k)}\\
&=\frac{e^{\epsilon}(1-P(A_{M}=1|S=i,\, EA_{M}=k))P(A_{M}=1|EA_{M}=k)}{1-P(A_{M}=1|EA_{M}=k)}.
\end{split}
\end{displaymath}
Rearranging the terms, we obtain the upper bound  $\Gamma(\epsilon,\delta_{k})$. 

Taking the derivatives of $\Gamma(\epsilon,\delta)$, we get $\frac{\partial\Gamma}{\partial\mathcal{\epsilon}}=\frac{e^\epsilon\delta(1-\delta)}{(1+\delta(e^{\epsilon}-1))^{2}} \geq 0$ and 
$\frac{\partial\Gamma}{\partial\delta}=\frac{e^{\epsilon}}{(1+\delta(e^{\epsilon}-1))^{2}}\geq 0$. Thus, $\Gamma(\epsilon,\delta)$ is increasing in $\epsilon$ and $\delta$. The inequality $\delta\leq \Gamma(\epsilon,\delta)$ follows from the observation that $\Gamma(\epsilon, \delta) - \delta= \frac{\delta (e^{\epsilon} - 1) (1 - \delta)}{1 + \delta  (e^{\epsilon} - 1)}\geq 0$, which holds since $1\geq\delta\geq 0$ and $e^{\epsilon}>1$. Finally, the limit as $\epsilon$ approaches zero can be computed straightforwardly as: $
\lim\limits_{\epsilon\rightarrow 0}\Gamma(\epsilon,\delta)=\frac{1\cdot \delta}{1+\delta\cdot 0}=\delta$.
\end{proof}

\begin{proof}[Proof of Proposition~\ref{prop:irr_strong}]
As in the proof of Proposition~\ref{prop:irr}, $P(C=1|S=i,\, EA=j)=\alpha_{i}\beta_{i,j}$ for any pair $(i,j)$. For the first part, observe: $\lambda<\alpha_{\mu'}\beta_{\mu',j'}\leq \alpha_{\mu'-1}\beta_{\mu'-1,j}$.
In particular, we have $\lambda<\alpha_{\mu'-1}\beta_{\mu'-1,j}$. This implies $\min\{i:P(C=1|S=i,\, EA=j)>\lambda\}\leq \mu'-1< \mu'$, which completes the proof of the first part. Similarly, for the second part, we have: $
\lambda<\alpha_{i'}\beta_{i',\kappa'}\leq\alpha_{i}\beta_{i,\kappa'-1}$. This implies $\min\{j:P(C=1|S=i,\, EA=j)>\lambda\}\leq \kappa'-1<\kappa'$, which completes the proof.
\end{proof}
\end{document}